\documentclass[peerreview]{IEEEtran}
\usepackage{cite}
\usepackage{amsmath,amssymb,amsfonts}
\usepackage{algorithm,algorithmic}
\usepackage{graphicx}
\usepackage{textcomp}
\usepackage{xcolor}
\usepackage[caption=false]{subfig}
\usepackage[final]{pdfpages}

\def\BibTeX{{\rm B\kern-.05em{\sc i\kern-.025em b}\kern-.08em
    T\kern-.1667em\lower.7ex\hbox{E}\kern-.125emX}}
\begin{document}
\title{Visual design intuition: Predicting dynamic properties of beams from raw cross-section images}
\author{
\IEEEauthorblockN{
    {Philippe M. Wyder\IEEEauthorrefmark{1}\IEEEauthorrefmark{2}, Hod Lipson\IEEEauthorrefmark{1}}\\ 
}
\IEEEauthorblockA{\tt\footnotesize\IEEEauthorrefmark{1}\textit{Department Of Mechanical Engineering}, 
        \textit{Columbia University}
        \textit{New York, USA}\\
} 
\IEEEauthorblockA{\tt\IEEEauthorrefmark{2} Corresponding Author: 
    \textit{philippe.wyder@columbia.edu}}
        
}
\maketitle
\thispagestyle{plain}
\pagestyle{plain}
\begin{abstract}
In this work we aim to mimic the human ability to acquire the intuition to estimate the performance of a design from visual inspection and experience alone. We study the ability of convolutional neural networks to predict static and dynamic properties of cantilever beams directly from their raw cross-section images. Using pixels as the only input, the resulting models learn to predict beam properties such as volume maximum deflection and eigenfrequencies with 4.54\% and 1.43\% Mean Average Percentage Error (MAPE) respectively, compared to the Finite Element Analysis (FEA) approach. Training these models doesn’t require prior knowledge of theory or relevant geometric properties, but rather relies solely on simulated or empirical data, thereby making predictions based on “experience” as opposed to theoretical knowledge. Since this approach is over 1000 times faster than FEA, it can be adopted to create surrogate models that could speed up the preliminary optimization studies where numerous consecutive evaluations of similar geometries are required. We suggest that this modeling approach would aid in addressing challenging optimization problems involving complex structures and physical phenomena for which theoretical models are unavailable. 
\end{abstract}
\begin{IEEEkeywords}
artificial design intuition, Shape optimization, biomimetics, deep learning, modal analysis, frequency analysis, surrogate model, geometric optimization
\end{IEEEkeywords}
\section{Goals and Motivations}
Arriving at an optimal design of a part, component, or structure often requires evaluating countless candidate solutions with similar geometries. This is commonly accomplished by performing finite element analysis (FEA) for each candidate design. This process is computationally expensive and time consuming, especially when the design problem involves complex geometries and complex nonlinear physics, and when sufficiently accurate theoretical models are unavailable. As finite element models cannot learn, but rather solve each problem anew, they cannot benefit from prior solutions of geometrically similar problems.\par
To overcome the computational burden of full FEA during optimization, engineers sometimes use surrogate models trained on sample cases to quickly predict approximate mechanical properties of candidate solutions without requiring a complete analysis. Once established, surrogate models can be used in the early exploratory phase of an optimization process.\par
Surrogate models are commonly created using machine learning (ML) techniques, allowing them to predict approximate structural properties when trained on examples from a relevant design domain\cite{Forrester2009, YAN2020105332, RAPONI2019730}. Typically, an ML algorithm receives the relevant physical and geometric design variables as input, and predicts the desired properties as outputs. For example, to predict beam deflection, the model would be provided with design variables such as beam length, material properties, and second moments. Using a dataset of input variables and corresponding outputs, an ML model can be trained to represent a function that approximates the resulting beam deflection.\par
Unfortunately, the relevant physical and geometric input variables for predicting a particular property are often not known in advance, especially when modeling complex phenomena, such as the energy absorption of a vehicle crash box, or the lift of a flexing insect wing. On the other hand, it is common knowledge that the second moment is a relevant input variable for predicting a beam's natural frequencies. In design spaces where both analytical models and knowledge of relevant design variables are lacking, surrogate models that learn the relevant variables would allow us to explore and optimize structures.\par
This was the motivation behind the present study, as a part of which an ML algorithm was adopted in an attempt to create a surrogate model from raw data inputs (such as image pixels) alone, i.e., without prior identification of the relevant input variables. If successful, the proposed strategy could be extended to tackle other problems for which the relevant input variables are unknown, such as those involving complex, nonlinear interactions: buckling under impact, aero-elastic vibrations, or the dynamic behavior of nano-reinforced micro structures with complex geometries. For example, the analysis of mechanical properties of micro-scale cantilever beams with carbon nanotube (CNT) reinforcements or structures with embedded CNTs or silica carbide tubes require FEA, since they are impossible to analyze with analytical methods \cite{Civalek2020, Civalek2021}. Thus, having a modeling approach that is able to identify the relevant parameters from a visual representation could facilitate the exploration of these design spaces where humans lack visual intuition.\par
Based on the results presented in this work, we believe that distilling ''visual design intuition'' through deep learning could lead to faster and more automated design optimization, especially for recurring problems that lack precise mathematical formulation.\par 
\subsection{Deep neural network models for visual intuition}
Deep learning methods, and in particular Convolutional Neural Networks (CNNs), have been successfully used in a wide range of design applications, from generating car silhouettes to predicting molecule structures \cite{Gunpinar2019,Chang, Carrigan2012, Liu2018}. 
Furthermore, CNNs have the capacity to model complex phenomena, such as classifying objects in images, recognizing actions in videos, and predicting the dynamic behavior of a lava lamp or a candle flame, as well as to predict the behavior of another agent in a game of hide and seek \cite{chen2020pixelphysics, chen200visualbehaviormodeling, chen200visualhideandseekphysical, chen2020visual}. 
However, in most cases, CNN models are trained on raw image data, without access to user-selected relevant variables or analytical models.\par
As humans, we have a naturally developed non-parametric visual intuition for our environment and the materials and structures that we typically interact with. This allows us to appreciate, for example, that a thinner beam has a lower natural frequency than a thicker beam of the same length, and that the tip of a longer beam will deflect more than that of a shorter beam with the same cross-section. Our intuition for shapes and materials similarly allows us to rule out flawed designs without relying on theoretical knowledge or analytical models.\par

\begin{figure}[htbp]
\centerline{\includegraphics[width=127mm]{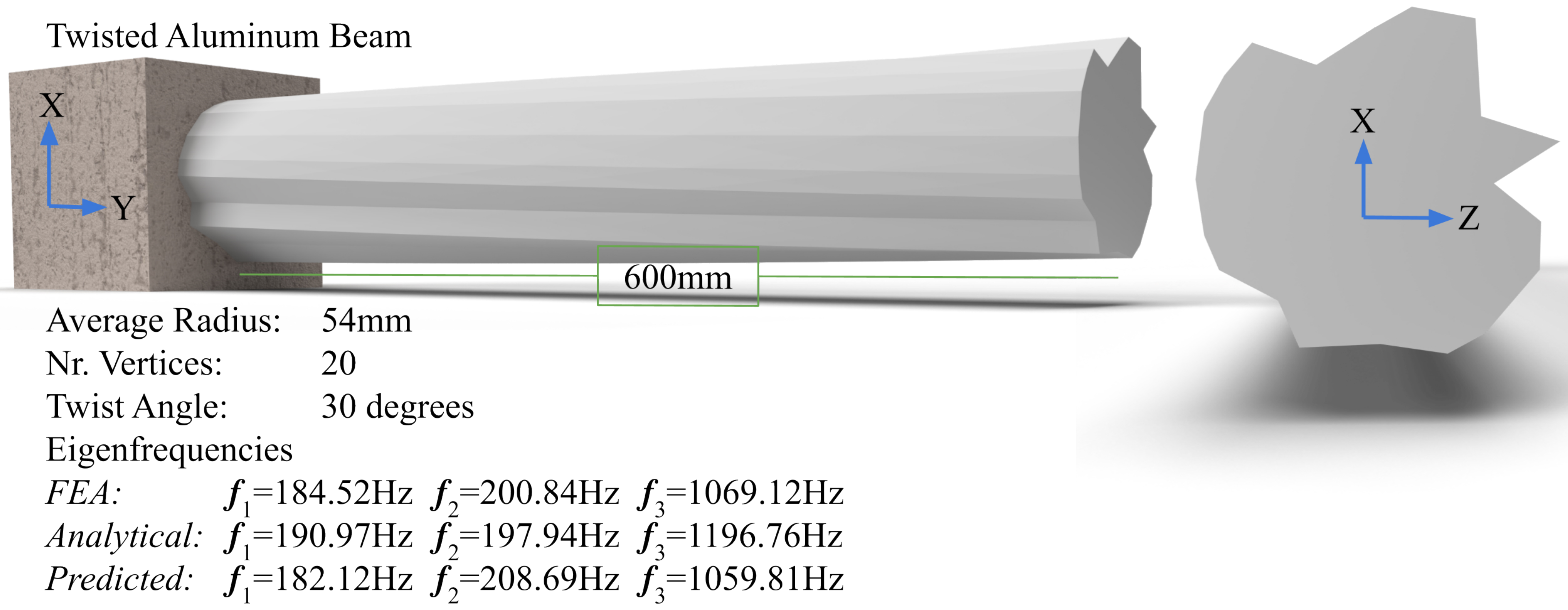}}
\caption{Twisted cantilever beam that is clamped at one end: the beam's first three eigenfrequencies ($f_1$, $f_2$, and $f_3$) are shown as they were estimated using FEA, an analytical solution, and our trained model. The FEA solution serves as the ground truth.}
\label{Fig1}
\end{figure}

Hence, the aim of the present study is to determine whether CNNs, when used as surrogate fitness measures for design optimization, can provide a similar visual design intuition. A trained engineer would struggle to guess the first eigenfrequency ($f_1$) of the beam shown in Fig. \ref{Fig1} solely from seeing its cross-section, even if provided with the information on the beam’s material and geometry. On the other hand, our proposed model that capitalizes on efficient GPU parallelization provided a correct answer to this particular question in approximately 2.3 ms with 1.61\% mean absolute error (based on the model presented in Appendix F-A). 
To achieve these results, we trained a CNN on cross-sectional images of randomly-generated cantilever beams to predict FEA outcomes, such as volume maximum stress, beam deflection, and eigenfrequency. The resulting model was subsequently applied as a surrogate fitness measure to identify beam geometries that fulfill predetermined configurations, such as a specific set of eigenfrequencies. \par
In sum, the model presented here was trained on previously computed FEA solutions and was effectively used as a surrogate fitness measure for design space exploration. To demonstrate the feasibility of our proposed approach, we developed the necessary modules to generate data, train a CNN model, predict mechanical properties, and generate designs for cantilever beams. As a proof of concept, we used our method to identify cantilever beams with a specific set of first three eigenfrequencies. The main contribution of this study stems from a novel method for analyzing and optimizing cantilever beams using a CNN that infers the beam’s mechanical properties from its cross-section. \par
\section{Background}
Engineers rely on computational methods such as FEM to optimize their designs. During the design process, analysis results guide engineers’ design decisions. This process can be enhanced through FEM as it partitions a complex problem (shape) into a mesh made up of many small sub-problems (shapes) for which solutions can be easily found, and an overarching solution pertaining to the whole shape can be derived. As each mesh element needs to be solved independently, the computational effort required is directly related to the number of mesh components and their respective degrees of freedom \cite{Okereke2018}. An adequately scaled mesh with the right mesh components is therefore crucial to finding an accurate solution within reasonable time. To facilitate meshing for common users, modern algorithms generate adaptive meshes that are refined in areas with more complex geometric features, such as curved surfaces or tight corners. Adaptive meshes reduce the computation time while improving the overall result accuracy. \par
Finding a design that exhibits specific physical behaviors, such as a beam geometry that resonates at a particular set of frequencies or deflects by a predetermined amount given a specific load and constraints, is a trial-and-error process and requires optimization algorithms to complete effectively. These algorithms require fitness measures to evaluate candidate solutions and guide their search. Consequently, optimization speed is mainly determined by the time it takes to evaluate the fitness of a design. Since, FEA requires extensive computational resources, it makes optimizations time intensive.\par
In our approach, this issue is overcome by supplementing FEA with a CNN model that learns from previously evaluated cases to guide future design exploration. The trained model provides an approximate quantitative measure for a design’s performance three orders of magnitude faster than an FEA analysis conducted using the same hardware. Akin to a skilled design engineer, our model learns from prior experience and gradually gains “design intuition” which it can apply when presented with a new design. \par
Given that training deep neural network models tends to be a data-intensive process, to demonstrate that our model does not suffer from this shortcoming, we provide evidence confirming that it yields reasonable levels of accuracy even when trained on a combined training and validation dataset only 80 data points in size (see section \ref{sec_data_efficiency}). Further, we show that our cross-section image resolution can be reduced by a factor of four while still producing excellent results (see Appendix D). Given the performance measures highlighted above and the current advances in the use of CNNs for modeling complex physical phenomena\cite{chen2020pixelphysics}, the approach presented in this work has the potential for use in a wide range of applications, especially those involving more complex geometries.\par
Neural networks can produce unexpected results when applied to areas for which training data is sparse or absent. Consequently, the engineering community has been justifiably cautious to adopt opaque data-driven models for critical applications, as false models could result in catastrophic outcomes. This risk can be mitigated by clearly defining the search space and ensuring that it is adequately spanned by the dataset. Additionally, it must be noted that models trained on FEA results are inherently less accurate than FEA. To mitigate such concerns, in Section \ref{sec_analytical_modelperformance}, we demonstrate the accuracy of our approach by comparing its performance with widely accepted analytical solutions. Finally, we propose trained models as a supplement to FEA. Our approach quickly provides an acceptable solution that can either be verified or perfected using numerical models.
\section{Literature Review}\label{sec_literaturereview}
Surrogate-based optimization has been extensively studied in engineering literature and has historically relied on classical machine learning (ML) methods rather than deep learning models. Yet, despite the adoption of more powerful ML methods, the implementation steps outlined by Forrester et al. have remained the same: (1) selection of variables to be optimized; (2) sampling the design space; (3) surrogate model selection and training; (4) exploration of the search space through the surrogate model; and (5) improvements to surrogate model based on the obtained results or verification of a satisfactory solution against the ground truth\cite{Forrester2009}. Our proposed approach is based on this same framework, albeit with changes to the surrogate model selection and data representation. \par
Authors of recent studies recognize the power of deep learning in the context of structural simulation. For example, Rizzo et al. predicted the critical flutter velocity of suspension bridges using an artificial neural network (ANN)\cite{Rizzo2020}. Similarly, Le et al. used an ANN as a surrogate model of fragility to improve the performance assessment of vertical structures under severe wind conditions\cite{Le2020}. In both cases, adoption of neural networks yielded significant performance improvements over numerical analysis.
Although CNNs are mainly used in image classification and segmentation, as well as natural language processing, they have been found to perform better than other types of ANNs in engineering applications.
For example, Oh et al. and Youqui Zhang et al. demonstrated that CNNs can be effectively employed for structural health monitoring using data in both time and frequency domain \cite{Oh2019,Zhang2019}.
CNNs can accept vectors of time series data as input in order to learn the relevant features, and then effectively make use of the inherent intertemporal patterns to output the predicted labels.\par
In their study, Finol et al. used both 1-dimensional and 2-dimensional CNNs to effectively model eigenvalue problems in mechanics\cite{Finol2019}. While an eigenvalue problem for the prediction of the first three eigenfrequencies of cantilever beams is also in focus of the present investigation, unlike Finol et al., we provide our CNN with only a gray-scale image of the beam cross-section, without predefining the relevant input variables.
As noted earlier, neural networks can be adopted to effectively model complex systems. However, training them requires considerable data resources, without making any inferences from known laws of physics. Karpatne et al. attempted to overcome these drawbacks by introducing physics-guided neural networks (PGNNs), which not only produce more physically accurate results, but also require less data to train\cite{Karpatne2017}. 
In a similar vein, Ruiyang Zhang et al. proposed a physics-guided convolutional neural network (PhyCNN) for seismic response modeling\cite{Zhang2020}. In both cases, the proposed models were better at generalizing while helping to counteract overfitting. Moreover, Ruiyang Zhang et al. showed that, by presenting frequency data as images in the frequency domain, these can serve as CNN input. Thus, they applied their PhyCNN as a surrogate model for structural response prediction. In this work, we show that even a less complex CNN, without incorporating laws of physics or sophisticated encodings, can be used as a surrogate fitness model for geometric optimization. 
As humans, we rely on our inherent ability to recognize shapes in our estimation of the physical properties of our environment. We mimic this capacity by using a bitmap image of the beam cross-section as our representation for our model input. A bitmap image can be kept constant in size while representing any shape independent of the number of vertices used to describe the cross-section.

\section{Materials and Methods}
The remainder of this article is organized into four sections: section \ref{subsec_datasetgen} Dataset Generation, section \ref{section_datasets} Datasets, section \ref{sec_neuralnetwork} Neural Network, and section \ref{subsec_optimization_algorithm} Optimization Algorithm. Thus, after presenting our data generation approach, we introduce the datasets used in our experiments. Next, we explain the different neural network configurations explored, as well as the architecture of the model carried forward. Finally, we address the adoption of a random search algorithm for geometric optimization of beam cross-sections in combination with our model as a surrogate fitness measure.
\subsection{Dataset Generation}\label{subsec_datasetgen}
Our data generation process comprises random beam generation, static analysis, and frequency analysis, for which we adopted the FreeCAD API to generate a Computer Aided Design (CAD) model of each random beam\cite{Riegel}, followed by static and eigenfrequency analysis of each beam using COMSOL Multiphysics 5.4\cite{COMSOL2018}. More detailed information on data generation, as well as our source code, can be found in Appendix H.\par
All beams considered in our simulations were assumed to be manufactured using aluminium with Young's modulus $E=70e^9N/mm^2$ and density $\rho=2.7e^{-6} kg/mm^3$. This was an arbitrary choice, given that the study aim was to demonstrate a CNN-based modeling approach, rather than examine behaviors of cantilever beams. As the goal of the analyses was to automatically process a range of beams, geometric non-linearity was enabled in the solver to increase the accuracy for outliers that may violate the geometric linearity assumption.\par
\subsubsection{Random Beam Generation}
Our investigation commenced with the generation of a random polygonal cross-section for each beam, which was then copied, scaled, rotated, and offset to create linearly extruded, tapered, and twisted beams using the loft command. The resulting beams were stored in the dataset as 3D CAD models, while the original cross-section was stored as a bitmap image.
When generating the cross-section shape, number of vertices, average radius, variance in angular spacing between vertices (irregularity), and variance in radii between vertices of the same shape (spikiness) were provided. The vertices of each cross-section in our datasets have a uniformly randomly sampled radius and number of vertices in the 24-63 mm and 3-30 range, respectively. The polar coordinate angle offsets and the radii of the polygon coordinates were sampled from a uniform distribution and a trimmed normal distribution, respectively. We adjusted the data generation process to minimize meshing errors during the automated FEA.
The complete cross-section generation algorithm can be found in Appendix B. 
\subsubsection{Static Analysis}\label{sec_static_analysis}
For each beam in the dataset, we performed static analysis on a clamped beam with a load at its tip. To automate the data analysis, we made use of the native COMSOL meshing functionality, which generates an adaptive tetrahedral mesh. We automated the static analysis for our datasets in our custom COMSOL app (see Appendix I), using the load cases presented in Table \ref{tbl_allDatasets}.
\subsubsection{Frequency Analysis}
Similar to the approach adopted for static analysis, when performing frequency analysis, we sequentially iterated over each beam in a custom COMSOL application (see Appendix I), with the beam clamped at one end (see fig. \ref{Fig1}). For each beam, we computed the first three eigenfrequencies and normalized participation factors (npf).\par
To avoid inflated eigenfrequency results, the mesh used for FEA needs to be as fine as possible. COMSOL Multiphysics offers nine preset mesh refinement levels, ranging from Level 1 (extremely fine) to Level 9 (coarse). Through repeated experiments, we found that Level 3 (finer) yielded the best trade-off between accuracy and computation time. We found that the computation time quadruples when the mesh level is reduced from 3 to 2, while the increase in accuracy is comparatively small.

\subsection{Datasets}\label{section_datasets}
Two datasets—SlenderBeamDS and TwistedBeamDS—were initially generated, respectively consisting of beams with linearly extruded cross-sections and beams that are both extruded and twisted. This distinction was made as the deflection and eigenfrequency of the beams comprising the SlenderBeamDS dataset can be determined analytically, whereas a common closed-form solution for twisted beams is not available.
To study the influence of beam geometry on model performance, we generated six additional datasets using the first 5,001 cross-sections stored in the TwistedBeamDS dataset, each of which contains beams with a different amount of twist and/or taper, as shown in Table \ref{tbl_allDatasets}. For example, TW30TA50\_DS contains beams with a $30^{\circ}$ twist and a 0.5 taper factor, i.e., the cross-section is scaled by 50\% over the length of the beam. Thus, beams within the same dataset differ from each other only in their cross-section geometry, as described in Appendix G.
\begin{table}[htbp]
\caption{Dataset specification table}
\begin{center}
\begin{tabular}{l|rrrrrrr}
\textbf{Dataset}               &
\rotatebox[origin=c]{55}{\textbf{\vtop{\hbox{\strut Linearly extruded}\hbox{\strut (SlenderBeamDS)}}}} &
\rotatebox[origin=c]{55}{\textbf{\vtop{\hbox{\strut 30 deg. twisted}\hbox{\strut (TwistedBeamDS)}}}} & 
\rotatebox[origin=c]{55}{\textbf{\vtop{\hbox{\strut Linearly extruded}\hbox{\strut (Linear\_DS)}}}} &
\rotatebox[origin=c]{55}{\textbf{\vtop{\hbox{\strut 50\% tapered}\hbox{\strut (TA50\_DS)}}}} &
\rotatebox[origin=c]{55}{\textbf{\vtop{\hbox{\strut 15 deg. twisted} \hbox{\strut (TW15\_DS)}}}} &
\rotatebox[origin=c]{55}{\textbf{\vtop{\hbox{\strut 15 deg. twisted} \hbox{\strut and 50\% tapered} \hbox{\strut (TW15TA50\_DS)}}}} &
\rotatebox[origin=c]{55}{\textbf{\vtop{\hbox{\strut 30 deg. twisted} \hbox{\strut and 50\% tapered}\hbox{\strut (TW30TA50\_DS)}}}} \\
\textbf{Size}                  & 17501                                    & 17501                               & 5001                           & 5001                         & 5001                         & 5001                             & 5001                             \\
\textbf{Twist Angle {[}deg{]}} & 0                                        & 30                                  & 0                              & 0                            & 15                           & 15                               & 30                               \\
\textbf{Taper Factor*}          & 1                                        & 1                                   & 1                              & 0.5                          & 1                            & 0.5                              & 0.5                              \\
\textbf{Irregularity}          & 0.4                                      & 0.4                                 & 0.4                            & 0.4                          & 0.4                          & 0.4                              & 0.4                              \\
\textbf{Spikiness}             & 0.1                                      & 0.15                                & 0.15                           & 0.15                         & 0.15                         & 0.15                             & 0.15                             \\
\textbf{Load Case {[}N{]}}     & {[}2000,0,0{]}                           & {[}1750,0,0{]}                      & {[}1750,0,0{]}                 & {[}1750,0,0{]}               & {[}1750,0,0{]}               & {[}1750,0,0{]}                   & {[}1750,0,0{]}                  \\
\multicolumn{8}{l}{* scale factor used to multiply the radii of the bottom cross-section vertices to obtain the top cross-section of the beam.}
\end{tabular}
\end{center}
\label{tbl_allDatasets}
\end{table}
\subsection{Neural Network}\label{sec_neuralnetwork}
\begin{figure}[ht]
\centerline{\includegraphics[width=180mm]{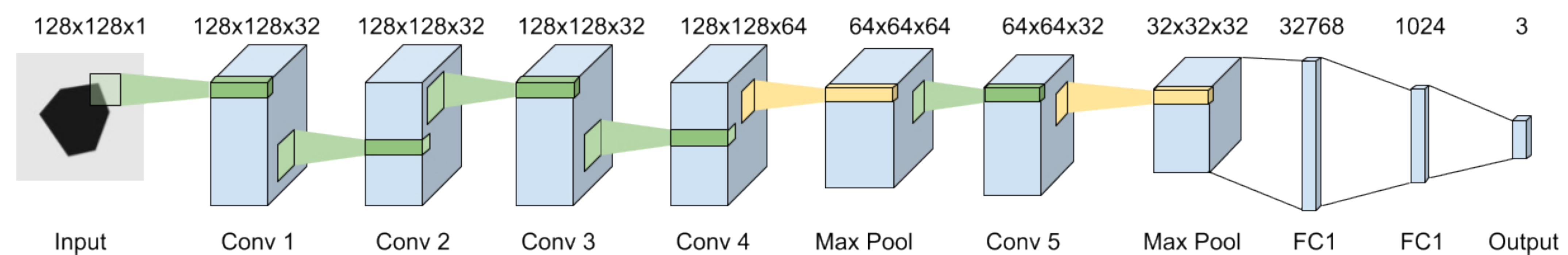}}
\caption{Neural network diagram of ConvNet Extended for predicting the first three eigenfrequencies of a cantilever beam.}
\label{fig_ConvNetExtended}
\end{figure}
As the objective of the present study is to propose a deep learning approach for predicting static and dynamic behaviors of cantilever beams from a single 2D cross-section image, the model is required to infer all other relevant variables, such as beam length or material properties. Thus, to limit the search space, all variables except the cross-section geometry were assumed to be constant for the entire dataset.
In the analyses, three neural networks—ConvNet (see Appendix A), ConvNet Extended, and Fully Connected (see Appendix A)—were explored. ConvNet and ConvNet Extended differ only in the number of convolutional layers. The Fully Connected network used in our tests consists of four fully connected layers.
We found that the ConvNet Extended (Fig. \ref{fig_ConvNetExtended}), the deepest CNN we tested, outperformed the other two architectures at predicting the first three eigenfrequencies (see Table I in Appendix A). The neural network was programmed in such a way that the output layer dimension changes if the network is trained to predict a different number of properties. For example, if the network were solely predicting the maximum beam deflection, the output layer would have one rather than three dimensions.\par
Unlike classical computer vision methods, CNNs learn the relevant features for predicting the output during training. Each layer is computed from the previous one by sliding a ''window'' called the kernel over the input tensor and applying the convolution operation. Moreover, as CNNs are able to extract learned features from data, they do not require any feature engineering, as described in detail by LeCun et al. and O'Shea et al. \cite{lecun1989backpropagation, oshea2015introduction}.\par
All our models were implemented in PyTorch and were trained and run on a desktop computer with Nvidia RTX2080Ti GPUs. Moreover, network size was adjusted based on the input image dimensions ($img\_size$) and the number of labels ($number\_of\_labels$) to be predicted.\par
Each dataset was segregated into training, evaluation, and testing subsets in a 64:16:20 ratio. For a dataset comprising 17,500 datapoints, this resulted in a 11,200:2,800:3,500 split. We trained each network using the Adam optimizer in combination with Mean Squared Error (MSE) (eqn. \ref{MSE_Equation}) as the loss function\cite{Kingma2015}. Unless specified otherwise, all models were trained with a learning rate (LR) of 0.0001.
\begin{IEEEeqnarray}{rCl}\label{MSE_Equation}\frac{1}{n}\sum_{i=1}^{n}(Y_i-\hat{Y_i})^2\IEEEyesnumber
\end{IEEEeqnarray}
\subsubsection{ConvNet Extended Architecture Details}
The \textit{ConvNet Extended} network comprises five convolutional layers and two fully connected layers and takes as an anti-aliased gray-scale image of the beam cross-section as input. The first three convolutional layers have 32 5×5 convolutions of stride 1 with padding 2, followed by batch normalization. The fourth convolutional layer has 64 5×5 convolutions of stride 1 with padding 2, followed by batch normalization and a max pool layer with a 2×2 convolution applied at stride 1. Finally, the fifth convolutional layer has 32 5×5 convolutions of stride 1 with padding 2, followed by batch normalization and a max pool layer with a 2×2 convolution applied at stride 1. All convolutional layers use ReLU activations\cite{Arora2018}. The sixth layer is a $32*img\_size/4*img\_size/4$ to $1024$ size fully connected layer and the final layer is a $1024$ to $number\_of\_labels$ fully connected layer.
\subsection{Optimization Algorithm}\label{subsec_optimization_algorithm}
To assess our model's effectiveness as a surrogate model, we used a trained model in combination with random search to identify beams that exhibit a specific set of first three eigenfrequencies ($[f_1, f_2, f_3]$) and confirmed our findings using FEA.\par
For the optimization, the best performing model (manually selected from multiple training sessions) was trained on the TwistedBeamDS dataset with an average MAPE of 2.03\%. The training graph of the model used in this experiment, as well as a table showing a detailed breakdown of the model's performance can be found in Appendix F-A.
During the optimization with the random search algorithm, we randomly generated new beam cross-section vertices using Algorithm 2 (see Appendix B). These vertices were subsequently used to create a cross-section image, as was done during data generation. After evaluating the resulting cross-section image using our model, we computed the MSE between the desired values for $[f_1, f_2, f_3]$ and the predicted values, saving the best-performing cross-section geometry after each iteration.
\section{Results}
The results yielded by our analyses are presented under the following five headings: Model Performance, Geometric Optimization of Beam Cross Sections, Data Efficiency, and Generalizability. In the section related to model performance, we show that our approach is not only superior to a closed-form solution, but can be applied to scenarios for which no readily available closed-form solution exists. The Optimization Algorithm section is designated for our experimental setup and the results yielded by applying our approach to identify specific beam configurations with desired properties. Next, in the Data Efficiency section, we discuss the data requirements of our model, and present the results obtained by training the model on various problems and datasets without tuning the relevant network variables in the Generalizability section.\par

\subsection{Model Performance}\label{sec_analytical_modelperformance}
To demonstrate the practical utility of the proposed approach, we compared the accuracy of our model to that of the analytical solution for both volume maximum beam deflection and eigenfrequency. As a closed-form solution for total volume maximum displacement and eigenfrequencies of a twisted cantilever beam is highly complex, the formula for linearly extruded beams was used in comparisons related to both linear and twisted beams\cite{Sinha2006}. Unlike the analytical solution, our finite element analysis was not constrained by the assumption of geometric linearity. As a result, larger discrepancies between the analytical and the finite element solution were expected for beams that do not satisfy the geometric linearity criterion.\par
\begin{IEEEeqnarray}{rCl}\label{eqn_principal_axes}\IEEEyesnumber
\theta_{p} = \frac{1}{2}tan^{-1}[\frac{2I_{xy}}{I_y-I_x}] \IEEEyessubnumber\label{eqn_principal_begin}\\
I_{x_p} = I_x cos^2\theta_{p}+I_y sin^2\theta_p-I_{xy} sin2\theta_p \IEEEyessubnumber\\
I_{y_p} = I_x sin^2\theta_{p}+I_y cos^2\theta_p+I_{xy} sin2\theta_p \IEEEyessubnumber\\
I_p=
\begin{bmatrix}
I_{x_p} & 0         & 0\\ 
0       & I_{y_p}   & 0\\ 
0       & 0         & 1
\end{bmatrix} \IEEEyessubnumber\label{eqn_principal_end} 
\end{IEEEeqnarray}
To analytically compute the maximum deflection and the first three eigenfrequencies of each beam, the second moment along the principal axes must be considered, which was computed with respect to the base coordinate frame during data generation. This allowed the equations \ref{eqn_principal_begin} through \ref{eqn_principal_end} to be applied to compute the second moment along the principal axes, as well as the z-axis rotation needed to transform the applied force vector into the principal axis coordinate frame.
\subsubsection{Total Deflection}
\begin{figure} 
    \centering
  \subfloat[\label{figX_a_TotDisp_SlenderBeamData}]{
       \includegraphics[width=0.55\linewidth]{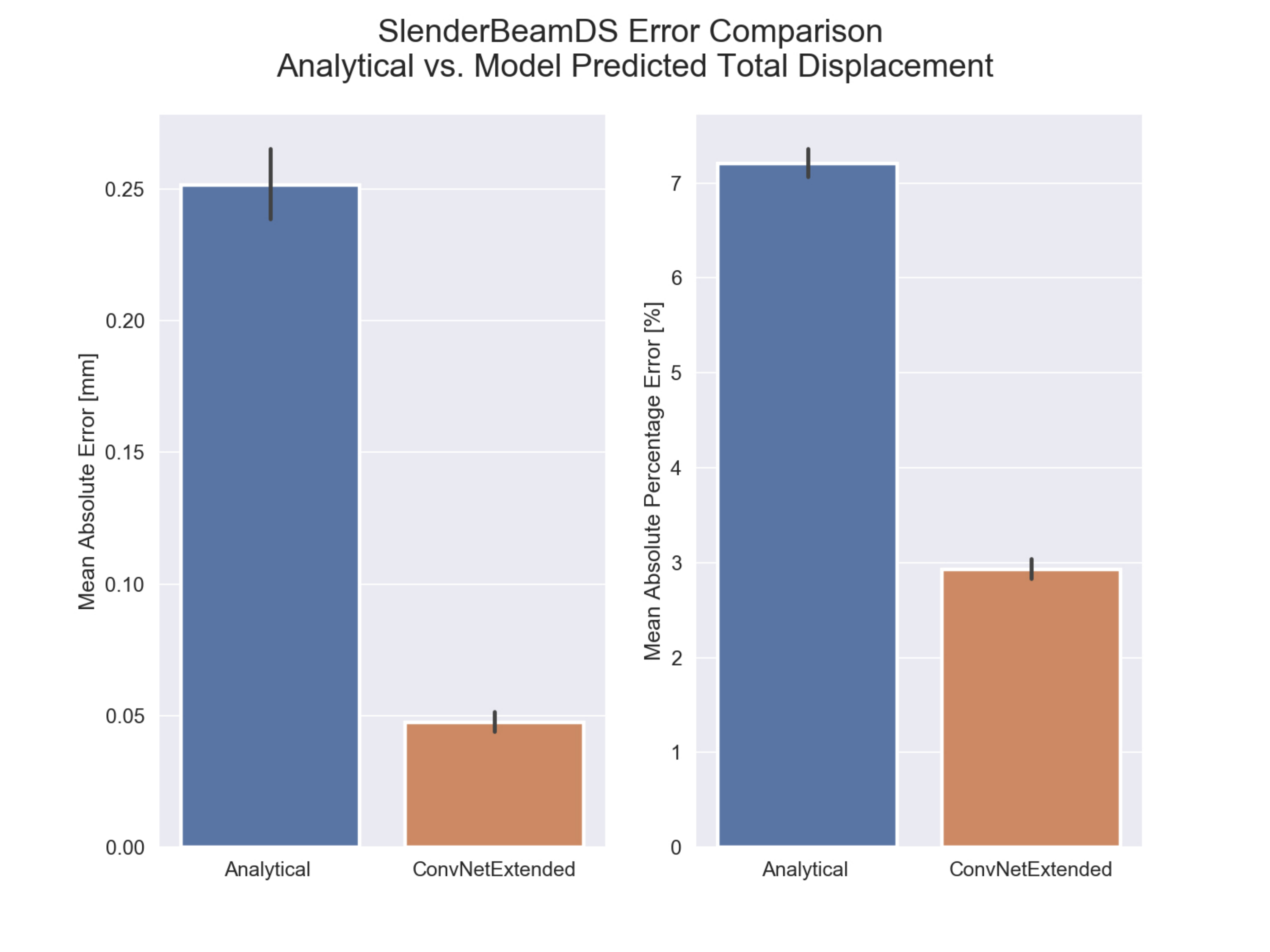}}
    \\
  \subfloat[\label{FigX_b_TotDisp_TwistedBeamData}]{
        \includegraphics[width=0.55\linewidth]{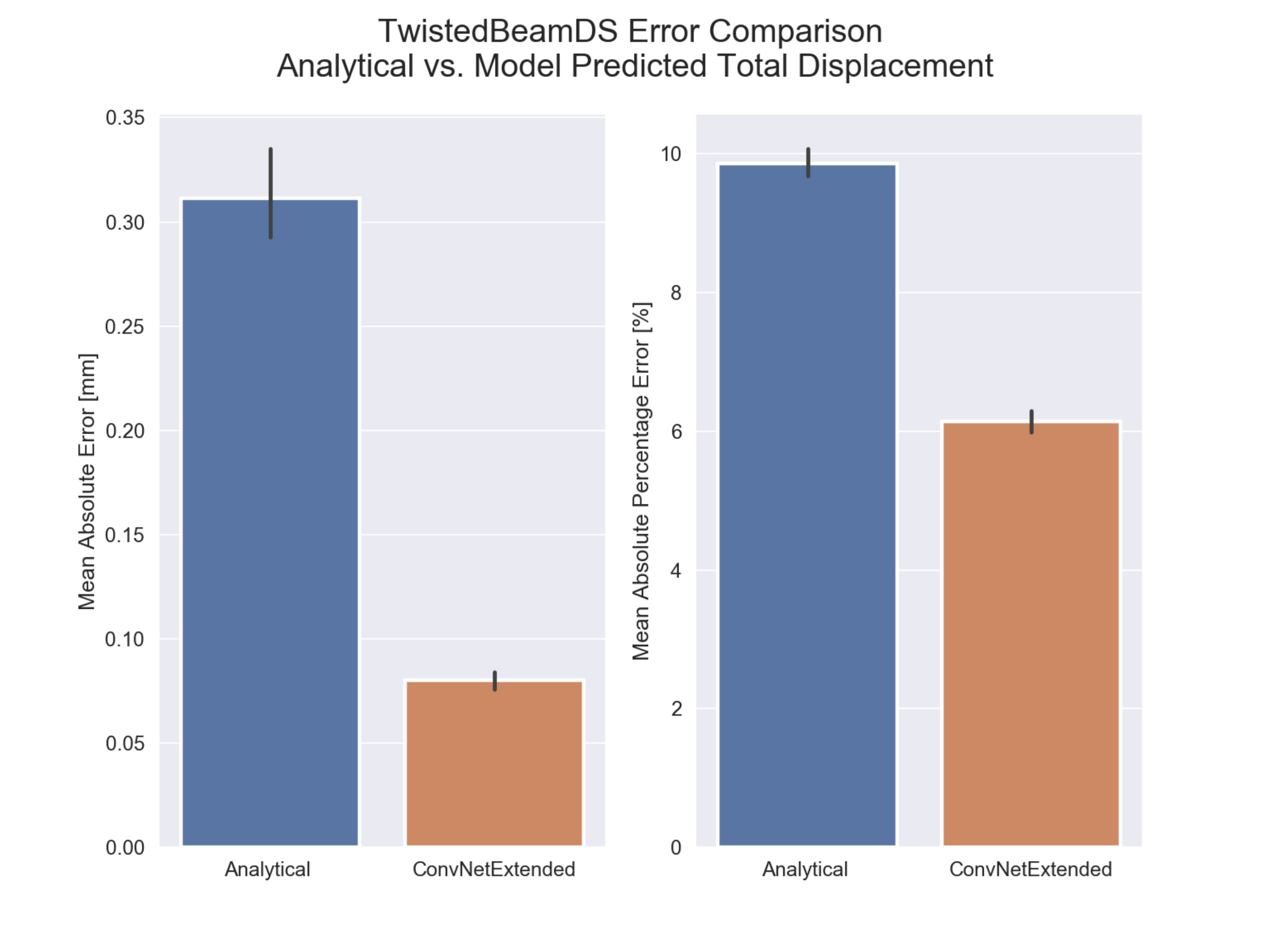}}
  \caption{Comparison of errors generated by the analytical solution (blue bar) and the model prediction (orange bar) of total displacement with respect to the ground truth FEA solution based on test data only, Fig. \ref{figX_a_TotDisp_SlenderBeamData} and \ref{FigX_b_TotDisp_TwistedBeamData} show the results related to the linearly extruded beams comprising the SlenderBeamDS dataset and the twisted beams included in the TwistedBeamDS dataset, respectively (see Table \ref{tbl_allDatasets} for dataset details). The error bars represent the 95\% confidence interval. The graphs confirm that our approach outperforms the analytical solution for free vibration in both scenarios.}
  \label{FigX_AnalticalVsPredicted_TotDisp} 
\end{figure}
We solved for the maximum deflection $\delta_{max}$ of a clamped cantilever beam subjected to a point load at its tip using equation \ref{eqn_beamDeflection}, where $F_p$ is the point force applied at the beam tip, $I_p$ is the second moment along the principal axis, and $E$ is the Young's modulus, as described in section \ref{section_datasets}. The load case considered in this scenario can be found in Table \ref{tbl_allDatasets}. A derivation of equation \ref{eqn_beamDeflection} is provided by Gere et al. in ''Mechanics Of Materials''\cite{Gere2009}. The point force $F_p$ was attained by representing the load case used in the FEA analysis in terms of the eigenbasis of the second moment.
\begin{IEEEeqnarray}{rCl}\delta_{max} = \frac{(F_pL)^{3}}{3EI_p}\IEEEyesnumber\label{eqn_beamDeflection} 
\end{IEEEeqnarray}
As shown in figure \ref{figX_a_TotDisp_SlenderBeamData}, our model is over four percentage points more accurate than the analytical solution for linearly extruded beams. Moreover, when applied to both linearly extruded and twisted beams, our model is approximately 0.2 mm more accurate in predicting volume maximum total displacement (see Fig. \ref{FigX_AnalticalVsPredicted_TotDisp}). 
One advantage of our approach stems from the assumptions imposed on the analytical solution related to the beam shape and material composition, leading to constraints such as linear bending. Thus, for beams that violate these assumptions, the analytical solution is naturally less accurate. Most importantly, our approach in combination with numerical approximations is a viable strategy for exploring the design spaces of complex problems that lack a known closed-form solution.
\subsubsection{First Three Eigenfrequencies ($[f_1, f_2, f_3]$)}
The eigenfrequency models used in this experiment were trained with 0.00001 learning rate and 100 batch size, which were determined based on the validation set performance.
\begin{figure}
    \centering
  \subfloat[\label{figX_a_ef123_SlenderBeamData}]{
       \includegraphics[width=0.55\linewidth]{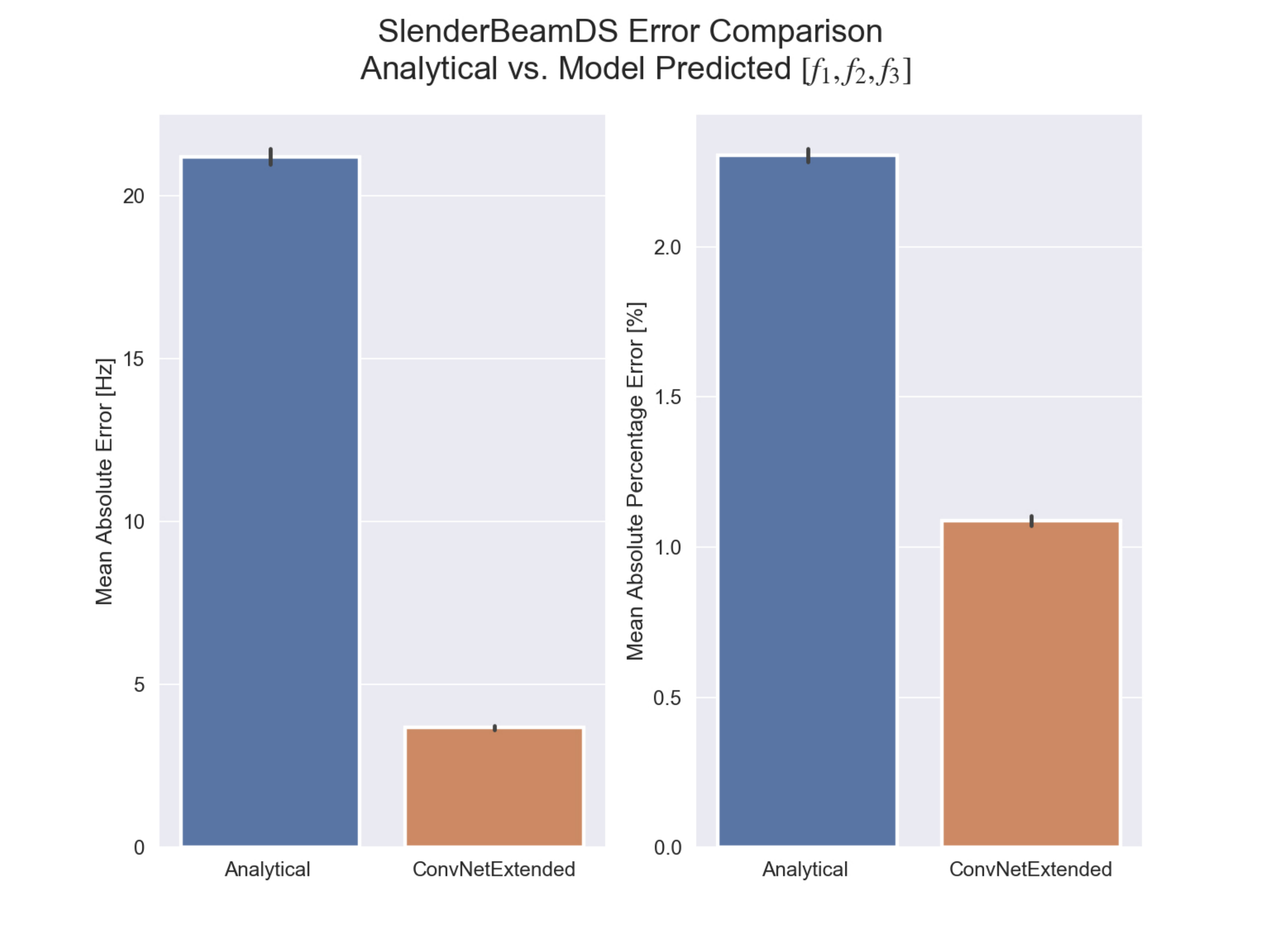}}
    \\
  \subfloat[\label{FigX_b_ef123_TwistedBeamData}]{
        \includegraphics[width=0.55\linewidth]{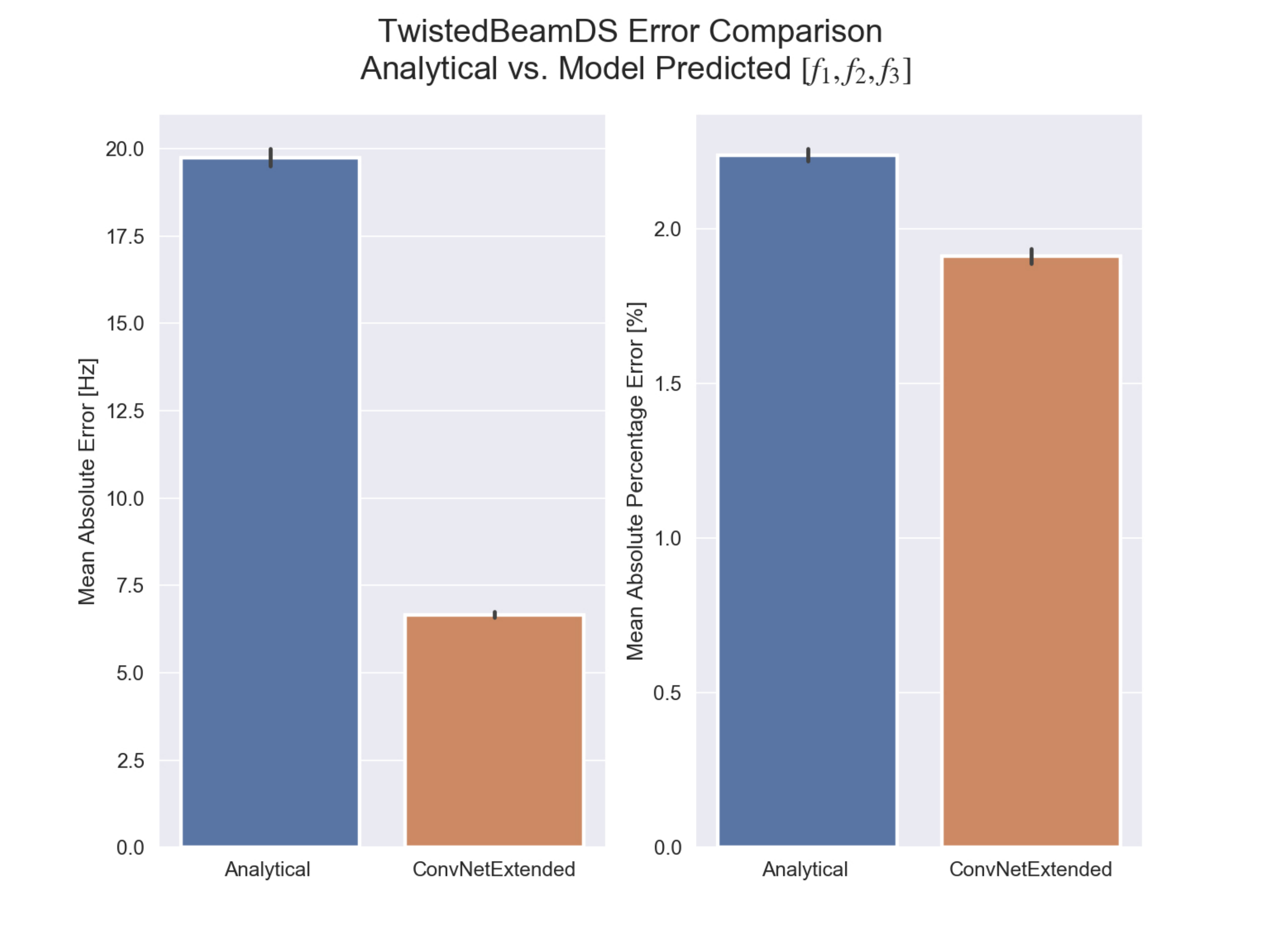}}
  \caption{Comparison of errors generated by the analytical solution (blue bar) and the model prediction (orange bar) of total displacement with respect to the ground truth FEA solution based on test data only, Fig. \ref{figX_a_ef123_SlenderBeamData} and \ref{FigX_b_ef123_TwistedBeamData} show the results yielded by applying our model to linearly extruded beams included in the SlenderBeamDS dataset and the twisted beams comprising the TwistedBeamDS dataset, respectively (see Table \ref{tbl_allDatasets} for dataset details). The error bars represent the 95\% confidence interval. The graphs show that our model outperforms the analytical solution for free vibration in both scenarios.}
  \label{FigX_AnalticalVsPredicted_EF123} 
\end{figure}
We next solved for the eigenfrequencies using the Euler-Bernoulli equations for free vibration of a cantilever beam using eqns. \ref{eqn_eigenfrequency_polynomial} to \ref{eqn_eigenfrequency_eigenfrequency} in combination with the root solutions provided in Table II in Appendix C. $\rho$ denotes material density, $A$ stands for cross section area, and $L$ is the beam length. For a detailed derivation please see Han et al. ''Dynamics of Transversely Vibrating Beams''\cite{Han1999}.
\begin{IEEEeqnarray}{rCl}\label{eqn_eigenfrequency}
\cosh{\beta_n}\cos{\beta_n}+1 = 0\IEEEyesnumber\IEEEyessubnumber\label{eqn_eigenfrequency_polynomial} 
\end{IEEEeqnarray}
\begin{IEEEeqnarray}{rCl}
\lambda_{m} = A\rho\IEEEyessubnumber\label{eqn_eigenfrequency_linearMassDensity} 
\end{IEEEeqnarray}
\begin{IEEEeqnarray}{rCl}
\omega_{n} = \frac{\beta_n^2}{L^2}\sqrt{\frac{EI_p}{\lambda_m}}\IEEEyessubnumber\label{eqn_eigenfrequency_eigenfrequency} 
\end{IEEEeqnarray}
\begin{IEEEeqnarray}{rCl}
f_n =\frac{\omega_{n}}{2\pi} \IEEEyessubnumber\label{eqn_eigenfrequency_eigenfrequency_fn} 
\end{IEEEeqnarray}
Our analyses revealed that the first three eigenmodes of the beams in our datasets occur along the two principal axes tangential to the beam length. Thus, we solved eqn. \ref{eqn_eigenfrequency_eigenfrequency} analytically in both directions and selected the lowest three eigenfrequencies among those found. It is worth noting that computing more than the first three eigenfrequencies may require solutions for all principal directions.\par
Our findings indicate that the proposed trained model, $ConvNetExtended$ outperforms the analytical solution for eigenfrequencies not only when applied to twisted beams, but also linearly extruded beams, as shown in figure \ref{FigX_AnalticalVsPredicted_EF123}. The trained model predicted the total displacement and the first three eigenfrequencies with half the percentage error of the analytical model when applied to linearly extruded beams (see Fig \ref{figX_a_TotDisp_SlenderBeamData} and \ref{figX_a_ef123_SlenderBeamData}). When applied to twisted beams, the trained model outperformed the analytical solution in terms of evaluating total displacement by over three percentage points (see Fig. \ref{FigX_b_TotDisp_TwistedBeamData}). When predicting the first three eigenfrequencies of twisted beams, the trained model outperformed the analytical solution by a small percentage margin, while being significantly more accurate with respect to the MAE (see Fig. \ref{FigX_b_ef123_TwistedBeamData}). In sum, our approach is not only more accurate than the analytical solution, but also more versatile, as it can be trained on diverse datasets and produces accurate results for various types of beam geometries that may not have an obvious closed-form solution, such as beams that are both tapered and twisted, as demonstrated in Section \ref{sec_generalizability}.
\subsection{Geometric Optimization Of Beam Cross Sections}
In this section, we study the performance of our models when employed as a surrogate fitness measure in the search for a beam cross-section geometry that exhibits a specific set of first three eigenfrequencies. 
For this purpose, we used a model that was trained to predict the first three eigenfrequencies based on the TwistedBeamDS dataset as our surrogate fitness model. Next, a random search algorithm was applied to identify cross-section designs that correspond to twisted beams with a desired set of first three eigenfrequencies. All geometries that met these criteria after a fixed number of evaluations were assessed via FEA and were used to compute the accuracy of our approach.\par
First, we randomly selected 100 sets of first three eigenfrequencies $[f_1, f_2, f_3]$ from beams in the TwistedBeamDS test set. As our model was trained on the TwistedBeamDS dataset, this strategy guaranteed that a beam that satisfies the desired eigenfrequency configuration exists, while excluding the possibility that the model had previously been presented with this particular set of eigenfrequencies. Second, we ran 10 optimization searches for each set of eigenfrequencies using a random search algorithm. Each search was terminated after 100,000 evaluations. Finally, the identified beam cross-sections were extruded into 3D CAD beams that were analyzed in COMSOL Multiphysics to determine their respective true first three eigenfrequencies.\par
\begin{figure}[ht]
\centerline{\includegraphics[width=95mm]{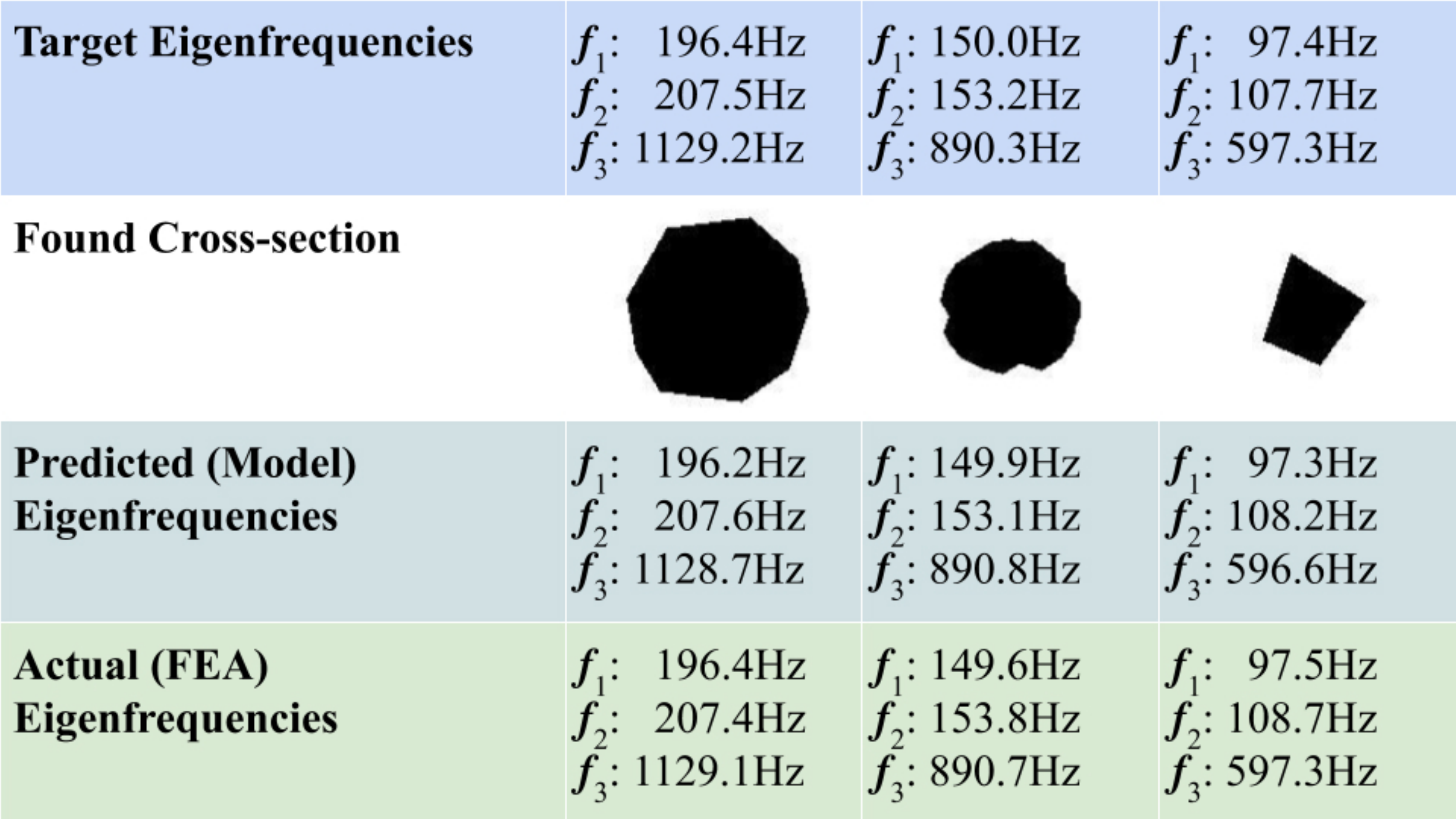}}
\caption{Three examples of the beam configurations yielded by the optimization experiment.}
\label{fig_FoundBeamsExamples}
\end{figure}
Our findings confirmed that the model developed as a part of this study was effective at guiding the random search algorithm toward beam cross-section geometries that provided a good fit to the target eigenfrequencies. The proposed solutions cumulatively deviated by 8.18 Hz on average over the three target eigenfrequencies resulting in 2.39\% MAPE. The identified cross-section geometries, once extruded with a 30 degree twist and analyzed using FEA, exhibited eigenfrequencies very close to the specified target. The three examples in Fig. \ref{fig_FoundBeamsExamples} show that the search produced cross-sections that differ in geometry, size, and vertex count to attain a beam geometry that best matches the target eigenfrequencies. The samples were manually selected from different frequency ranges to provide insight into the pertinent cross-section geometries.\par
We  found that on average meshing and analyzing one beam in COMSOL Multiphysics takes 4.3 seconds on an AMD Ryzen Threadripper 2950X CPU using meshing refinement level 3, while evaluating a new beam cross-section using our model on an Nvidia RTX2080Ti GPU was completed within 2.3 milliseconds. As a result, using FEA instead of our trained model for this parameter search would extend the 4-minute duration of each 100,000-step optimization to a week. Additionally, we were able to shorten the time required for the 100 optimization runs further by running seven surrogate model instances in parallel on one Nvidia RTX2080Ti. The time comparison between GPU parallelized code and code that runs on the CPU is not an even one, but is used in this instance to show the significant performance difference of the two methods on desktop computer hardware.\par
Analyzing the frequency response and tuning the components to resonate or not resonate at particular frequencies is relevant in many practical applications, from civil engineering to space travel. For example, our approach can be utilized when working on a mechanical resonator design, as it can identify geometries that vibrate at specific frequencies. Similarly, this approach can speedup the search for geometries that avoid specific eigenfrequencies, for example when designing machine parts that must avoid resonance with motor vibrations. 
\subsection{Data Efficiency: Prediction accuracy as a function of dataset size}\label{sec_data_efficiency}
Thus far, we assessed the performance of our approach using a large dataset. However, for design intuition to be useful, it should be adequately trained with a reasonably small amount of data. In a surrogate-based optimization approach as outlined in section \ref{sec_literaturereview}, the data quantity required to train the model determines its practical utility. Thus, in this section, we show that our model can be trained effectively on relatively small datasets without compromising its high prediction accuracy.\par
\begin{figure}[htbp]
\centerline{\includegraphics[width=95mm]{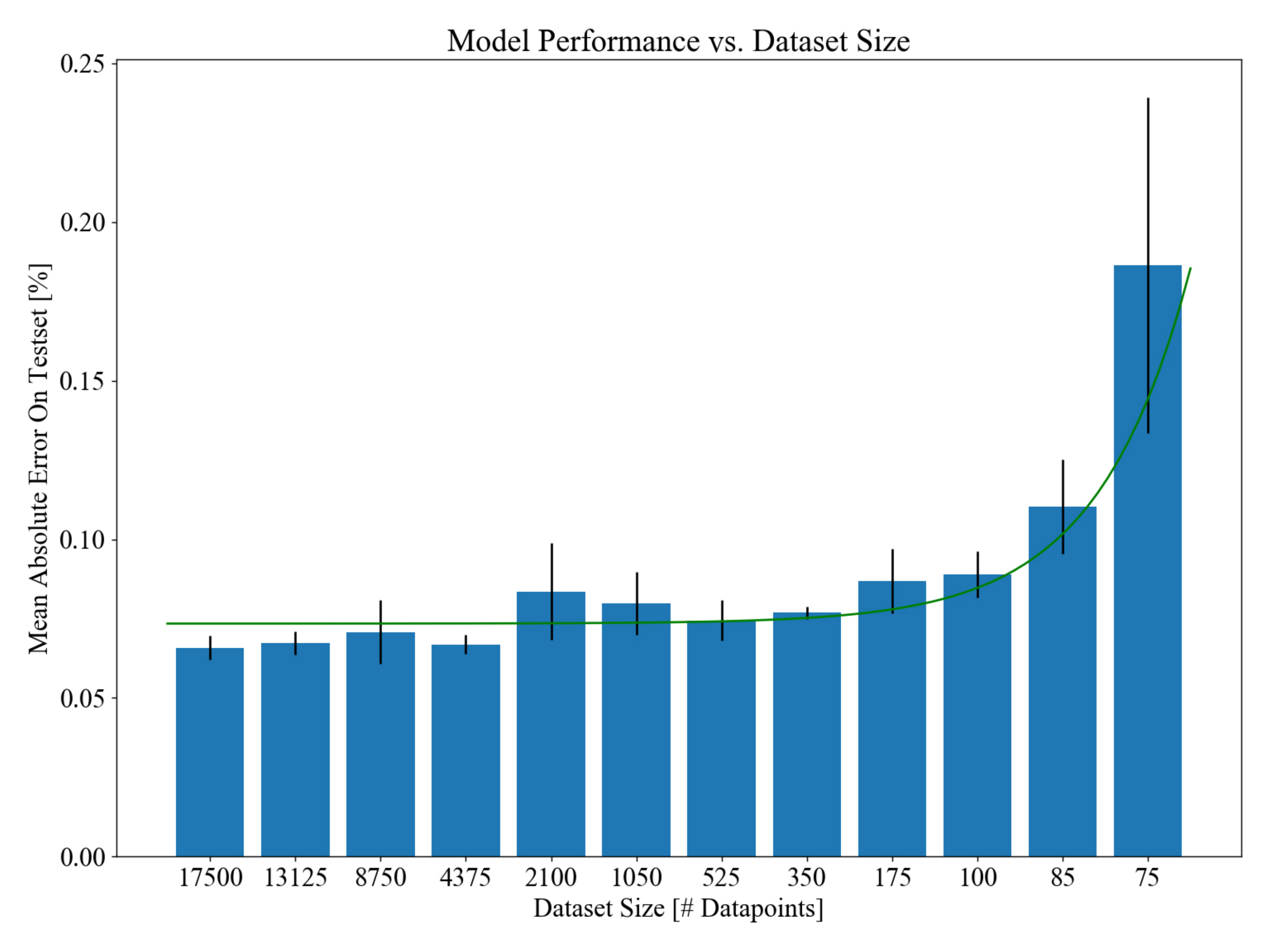}}
\caption{Barplot of the model's mean absolute percentage error versus the dataset size used to train the model.}
\label{fig_8_DataEfficiencyPlot}
\end{figure}
For this purpose, we partitioned the TwistedBeamDS dataset into differently sized subsets, and trained four models on each subset to predict the first three eigenfrequencies. We reserved 20\% (i.e., 3,500 data points) of the TwistedBeamDS for testing, and computed the mean average percentage error for each model based on the models’ performance on that test set. Each subset was split in 64:16:20 ratio: 64\% of the data was used for training, 16\% for validation, and the remaining 20\% was omitted, since we used the previously selected 3,500 data point test set for testing. The Dataset Size in Figure \ref{fig_8_DataEfficiencyPlot} refers to the total number of data points in the subset; assuming a 64:16:20 split, for a subset consisting of 100 samples, 80 would be designated for training and validation.\par
The obtained results are presented in Figure \ref{fig_8_DataEfficiencyPlot}. As indicated by the green trend line, the models trained on a larger dataset generally outperformed those trained on less data. However, it is also apparent that our approach performed less than 2.3 percentage points worse with 8.89\% MAPE when trained on 80 data points relative to the full dataset comprising 17,500 data points. These findings confirm that even a small initial dataset is sufficient to provide intuition about the design space, and can be a useful starting point for further design exploration.

\subsection{Generalizability: Model performance on different labels and beams types}\label{sec_generalizability}
To determine whether our approach can be applied to model various beam geometries, we trained three models on every combination of 12 labels and 7 datasets and reported the results in Table III in Appendix E. Based on our previous tests, we found that our approach performs well even if the input image is of low resolution (see Appendix D). For this evaluation, we trained the models using images with 64×64 pixel resolution to reduce the memory requirements and the overall computation time. The results show the average MAPE and the standard deviation for all three models based on all dataset/label combinations.\par
Overall, our approach achieved an average of 4.07\% MAPE, considering the best performing model for each dataset and label combination. This high level of accuracy was attained without tuning the training-relevant variables of the CNN for the respective labels. This is a particularly impressive achievement considering that it is not possible to deduce from a cross-section image whether a beam is twisted or tapered, or both. Twisted beams tend to bend out of plane, and thus produce different TotDisp results. Similarly, tapered beams exhibit a different frequency profile from linearly extruded beams due to the difference in mass and shifted center of mass. Hence, these results confirm that our approach can accurately predict both static and dynamic behaviors of linearly extruded, tapered, twisted, and tapered and twisted cantilever beams.\par
We also found that our model's performance suffers when trained on non-normalized labels that span a wide range of values. This is not unusual for machine learning models and can be addressed by scaling or normalizing the model inputs and outputs. To mitigate this issue, we scaled the values for curl displacement in the y-direction (CurlDisp\_Y) and principal strain in the x-direction (PrincStrain\_X) and achieved performance comparable to that obtained for other labels (see Appendix E). Moreover, even though our model's performance on volume maximum total displacement (TotDisp) and total Von Mises stress (TotVonMises) was sufficiently accurate, careful input and output scaling or normalization would improve the accuracy even further.\par
Frequency analysis helps engineers identify eigenfrequencies with significant mass participation to make design decisions or to inform further analysis. Hence, identifying a quantitative measure of an eigenfrequency's impact on the analyzed structure can inform further analyses. For this purpose, we included the root mean squared (RMS) of the normalized participation factor (npf1\_RMS to npf3\_RMS) as a label in our analysis as a proxy measure for the mass participation of each predicted eigenfrequency. Our model successfully predicted the root mean squared (RMS) of the normalized participation factor with an average MAPE of 2.11\%.
\section{Challenges and Limitations}\label{ChallengesAndLimitations}
We focused our study on eigenfrequencies and maximum deflection to ascertain if our approach is capable of predicting both static and dynamic behaviors of different beam geometries. While we recognize that eigenfrequencies are rarely used in practice in the absence of their corresponding mode shapes, their exploration is beyond the current study, as it requires a neural network architecture that differs significantly from the one presented in this paper.
 
Since our proposed model can’t be more accurate than the data it was trained on, it is important that the right type of FEA is performed for the problem at hand. At the micro and nano scale, cantilever beams behave differently than at the macro scale \cite{Demir2017, Numanoglu2018}. Therefore, the correct analysis approach must be used to get accurate data to train a surrogate model and verify optimization results.

Further, many situations require engineers to study more than the first three eigenfrequencies. Although this scenario was not explored in this work, it is likely that the proposed approach is capable of predicting a larger number of eigenfrequencies without significant changes to the model.
 
Finally, based on the results presented in this paper, it cannot be speculated whether this approach generalizes across other applications, search spaces, or geometries. Nonetheless, as it is capable of predicting different beam properties from a variety of beam geometries it has the potential for use in a wide range applications. 
\section{Conclusion}
We have demonstrated that finite element analysis results can be used to impart visual design intuition to CNNs, and that the resulting models can be used to optimize the shape of a design and to search for specific physical properties. Our model effectively learned implicit physical and geometric characteristics, such as beam length, moment of inertia, and material properties. Owing to its data efficiency, it can be incorporated into hybrid surrogate optimization approaches where a search space is randomly sampled to train a model that subsequently serves as a surrogate fitness measure to find better design solutions, which are numerically verified before being included into the dataset to improve the model.\par
We believe that adoption of our approach can significantly improve productivity in work environments where FEA is frequently performed on similar shapes, or a corpus of design-relevant solved analyses already exists. In such cases, the data pertaining to solved simulations can be used to train CNNs to expedite the work on future design optimization problems. A variety of cloud platforms already provide FEA capabilities to their customers, and have the data to capitalize on this method. By capturing each analysis and the corresponding results, deeper and more complex models with broader applications could be trained and used to guide design explorations.\par
Most importantly, our approach can be applied to poorly understood processes, as it is capable of implicitly learning to extract and compute the relevant physical and geometric variables from the raw visual data provided, without the need for manual feature engineering. Owing to its simplicity, it can serve not only as a means for furthering design optimization in established fields, but also as a tool for exploring poorly understood yet important new design spaces. Our approach could create models that mimic expert intuition in areas where we lack expertise.\par 
\section{Future Outlook}
While acknowledging that the work presented in this paper had a very narrow focus—cantilever beams—the models employed are capable of implicitly learning relevant variables, such as material, beam length, twist angle, etc. Going forward, we plan to design deep learning models that accept these design-relevant parameters as inputs to inform their predictions. We also aim to improve our approach and increase its ability to generalize by exploring alternative neural network architectures or using conditional neural networks. We believe that supplementing FEA with CNNs will accelerate and expand our ability to explore search spaces and optimize designs.\par
Ultimately, we would like to apply this approach to optimizing structures with multiple-members. These are the six steps that we believe are needed to employ the approach demonstrated in this paper to optimize more complex structures: (1) Identify the design parameters that will be optimized, and define the multi-member structure in terms of it. (2) Find an appropriate representation that can be used as input for the model. While a cross-section image was sufficient for our application, a more complex structure may require multiple images, a 3D voxel representation, or an alternative representation all together. (3) Create a high-fidelity simulation. (4) Generate an initial dataset that spans the desired design space, by randomly sampling the design space around the area of interest. (5) Train the CNN surrogate model on the generated dataset. (6) Validate the found design solution against the high-fidelity simulation. If the model's predicted solution does not accurately match the simulated solution, then sample more data points at and near the found configuration, add those data points to the generated dataset, and return to step (5).\par
\section{Acknowledgements}
We thank Joni Mici and Robert Kwiatkowski for reviewing our work and providing feedback on this manuscript. We also thank Michael Ounsworth for sharing his implementation of a polar coordinate-based polygon generation algorithm, which we integrated into our data generation process: see https://stackoverflow.com/questions/8997099/algorithm-to-generate-random-2d-polygon.

\section{Author Contributions}
Philippe Wyder - Experiments and Manuscript\par
Hod Lipson - Conceptualization and Editing
\section{Data Accessibility}
The appendix to this document is in the Supplementary Materials PDF.
Link to data generation repository: https://github.com/ResearchMedia/CantileverBeamDatasetGenerator\par
Link to neural network repository: https://github.com/ResearchMedia/CantileverNN\par
The data was published published on Mendeley Data. Doi: 10.17632/y3m8xm6kfk.1\par

\section{Funding Statement}
This work has been funded in part by U.S. Defense Advanced Research Project Agency (DARPA) TRADES grant number HR0011-17-2-0014.\par
This work was supported in part by the U.S. National Science Foundation (NSF) AI Institute for Dynamical Systems (dynamicsai.org), grant 2112085\par
The authors declare that they have no other known competing financial interests or personal relationships which have, or could be perceived to have, influenced the work reported in this article.

\bibliographystyle{acm}




\section*{Supplementary material}

# Visual design intuition: Predicting dynamic properties of beams from raw cross-section images
## *Supplementary Materials*

Philippe M. Wyder*[†], Hod Lipson*
*Department Of Mechanical Engineering, Columbia University New York, USA
[†] Corresponding Author: philippe.wyder@columbia.edu

## APPENDIX A
### NEURAL NETWORK ARCHITECTURES

In this section, we provide additional information on the network architectures that we explored before choosing ConvNet Extended. We present the performance achieved using each architecture in our test in table I. Further, we providing some notable observations on those results.

TABLE I
NETWORK ARCHITECTURE PERFORMANCE COMPARISON

| Exp. | Network | MAE [Hz] | MAPE [%] | LR |
|---|---|---|---|---|
| 1 | Fully Connected | 33.9 | 10.6 | 0.0001 |
| 2 | ***ConvNet Extended*** | ***20.5*** | ***6.1*** | ***0.0001*** |
| 3 | ConvNet | 24 | 7 | 0.0001 |
| 4 | Fully Connected | 22.5 | 6.7 | 0.00001 |
| 5 | ConvNet Extended | 20.8 | 6.1 | 0.00001 |
| 6 | ConvNet | 21.5 | 6.6 | 0.00001 |

*1) ConvNet:* ConvNet consists of two convolutional and two fully connected layers. The usual input to the network is a 128×128×1 anti-aliased gray-scale image of the beam cross-section. The first convolutional layer has 64 5×5 convolutions of stride 1 with padding 2, followed by batch normalization. The second convolutional layer has 32 5×5 convolutions of stride 1 with padding 2, followed by batch normalization and a max pool layer with a 2×2 convolution applied at stride 1. Both use ReLU activations [1]. The third layer is a $32 * img\_size/2 * img\_size/2$ to 1024 size fully connected layer and the fourth layer is a 1024 to $number\_of\_labels$ fully connected layer.

*2) Fully Connected Network:* The fully connected network used in our experiment is four layers deep, and its usual input is a 128×128×1 gray-scale image. All four layers are fully connected and have the input size of 65k, 16k, 16k and 1024, with 16k, 16k, 1024 and $number\_of\_labels$ output size, respectively.

### A. Observations on neural network architectures

Based on the findings reported in table I, we adopted ConvNet Extended for all subsequent experiments. We observed that, when applied to our beam datasets, the Fully Connected Network performed on par with the ConvNet and was in some cases superior. We ascribe these results to the larger size of the fully connected network used in our test. Although the Fully Connected Network had the same layer count as our smaller ConvNet, it has an order of magnitude more connections than even the ConvNet Extended model.

## APPENDIX B
### POLYGON GENERATION ALGORITHM

The randomly chosen cross-sections that were used to extrude the 3D beams were generated using algorithm 1 and 2. Our implementation is a close adaptation of Michael Ounsworth's code excerpt shared on the Stack Overflow forum (https://stackoverflow.com/a/25276331).

## APPENDIX C
### ROOTS FOR ANALYTICAL EIGENFREQUENCY CALCULATION

The roots used for the analytical eigenfrequency calculations are shown in table II.

**Algorithm 1** clip
**Input:** x, min, max
**Output:** clippedValue
 1: **if** ($min > max$) **then**
 2:    **return** x
 3: **else if** ($x < min$) **then**
 4:    **return** min
 5: **else if** ($x > max$) **then**
 6:    **return** max
 7: **else**
 8:    **return** x
 9: **end if**

---

**Algorithm 2** Polygonal cross-section vertices generator
**Input:** ctrX, ctrY, aveRadius, irregularity, spikiness, numVerts
**Output:** vertices
   *Initialisation* :
 1: vertices = []
 2: angularVariance = $clip(irregularity, 0, 1) * 2 * \pi / numVerts$
 3: radiusVariance = $clip(spikiness, 0, 1) * aveRadius$
   *Sample Angle Steps:*
 4: angleSteps = []
 5: minTheta = $(2 * \pi / numVerts) - angularVariance$
 6: maxTheta = $(2 * \pi / numVerts) + angularVariance$
 7: sum = 0
 8: **for** $i = 0$ to $numVerts$ **do**
 9:    tmp = $Uniform(lower, upper)$
10:    angleSteps.append( tmp )
11:    sum = $sum + tmp$
12: **end for**
   *Normalize vertex angles:*
13: k = sum * $2 * \pi$
14: **for** $i = 0$ to $numVerts$ **do**
15:    angleSteps[i] = $angleSteps[i]/sum$
16: **end for**
   *Sample radii and generate Cartesian coordinates:*
17: angle = $Uniform(0, 2 * \pi)$
18: **for** $i = 0$ to $numVerts$ **do**
19:    $r_i = clip(Normal(aveRadius, spikiness), 0, aveRadius)$
20:    $x_i = ctrX + r\_i * cos(angle)$
21:    $y_i = ctrY + r\_i * sin(angle)$
22:    vertices.append( (int($x_i$), int($y_i$)) )
23:    angle = angle + angleSteps[i]
24: **end for**
25: **return** $vertices$

## APPENDIX D
### THE EFFECT OF INPUT IMAGE RESOLUTION ON PREDICTION ACCURACY

We explored the effect of using lower-resolution anti-aliased cross-section images to train our model to predict the first three eigenfrequencies. As discussed in section IV-C, our models were programmed to adjust to their input image size. We generated anti-aliased gray-scale cross-section images with 32×32, 48×48, 64×64, 96×96, and 128×128 pixel resolutions. We trained each network four times, adopting 0.001, 0.0001, 0.00001, and 0.000001 as the learning rate, respectively, in an effort to reduce a performance bias due to a single choice of learning rate.

As shown in the box plot in fig. 1, model performance was high at all resolutions, as indicated by an average mean absolute percentage error of below 2.5%. Surprisingly, a larger standard deviation was obtained for images with higher resolution, and in these cases models performed slightly worse than when applied to images with lower resolution. These findings indicated

1

TABLE II
ROOTS $\beta_n$ OF EQN. 4A IN ASCENDING ORDER

| $n$ | $\beta_n$ |
|---|---|
| 1 | 1.8751040687119611664453082410782141625701117335 3107... |
| 2 | 4.6940911329741745764363917780198120493898967375 4577... |
| 3 | 7.8547574382376125648610085827645704578485419292 3005... |
| 4 | 10.995540734875466990667349107854702939612972774 6516... |

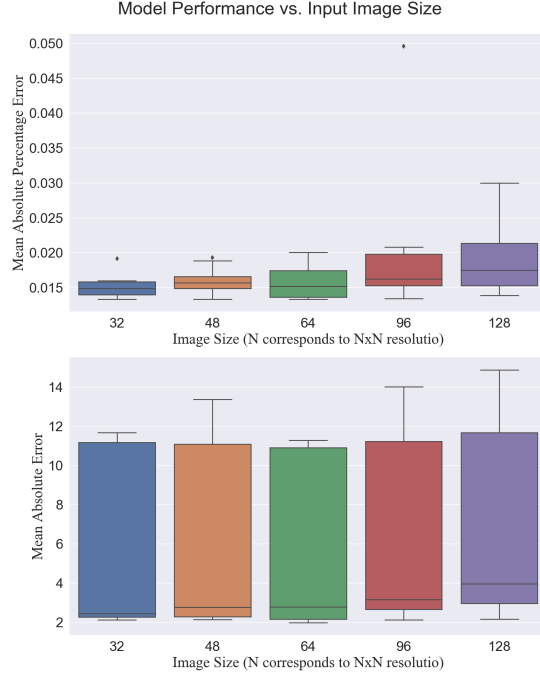

Fig. 1. This figure shows the effect of training our model on input images of different sizes. It is evident that the input image size did not significantly affect model performance. Interestingly, some models trained with smaller images outperformed other models that were trained on larger input images.

that the resolution of our cross-section representation was over-refined. We encoded $1mm^2$ per pixel in our 128×128 pixel image, and could potentially decrease the encoding resolution to $4mm^2$ per pixel given that the model trained on a 32×32 pixel image exhibited comparably good performance. This would also allow the same encoding to be applied to a wider beam size range. Our model's performance did not deteriorate even when trained on significantly less data in terms of pixel count, and it was able to extract salient information even from a smaller image.

## APPENDIX E
### DATASET VS. LABEL PERFORMANCE TABLE

Table III shows the model performance when predicting twelve labels for seven datasets. The first four labels correspond to static analysis results: volume maximum total displacement (TotDisp), total Von Mises stress(TotVonMises), curl displacement in the y-direction (CurlDisp_Y) scaled by a factor of 1,000, and principal strain in the x-direction (PrincStrain_X) scaled by $1e6$. The remaining eight labels pertain to the frequency analysis results: the first ($f_1$) through third eigenfrequency ($f_3$); the first three eigenfrequencies together ($[f_1, f_2, f_3]$); the RMS of the normalized participation factors of the first three eigenfrequencies (npf1_RMS, npf2_RMS, npf3_RMS), and the RMS of the normalized participation factors of the first, second, and third eigenfrequencies combined (npf123_RMS). The RMS of the normalized participation factors is a proxy measure for the mass participation factor corresponding to each eigenfrequency. The RMS of the normalized participation factor allows us to predict a single value rather than six for each eigenfrequency, while providing a sense of each frequency's significance. We trained our model to predict multiple beam properties simultaneously, for the first three eigenfrequencies, as well as for the RMS normalized participation factors corresponding to the first three eigenfrequencies. All relevant details on the datasets used are reported in table I.

TABLE III
MEAN AVERAGE PERCENTAGE ERROR AND STANDARD DEVIATION OF OUR MODEL TRAINED THREE TIMES ON DIFFERENT DATASET AND LABEL COMBINATIONS

| Label / Dataset | Linearly extruded (Linear_DS) | Linearly extruded (SlenderBeamDS) | 50% tapered (TA50_DS) | 15 deg. twisted and 50% tapered (TW15TA50_DS) | 15 deg. twisted (TW15_DS) | 30 deg. twisted and 50% tapered (TW30TA50_DS) | 30 deg. twisted (TwistedBeamDS) | Label Description |
|---|---|---|---|---|---|---|---|---|
| **TotDisp** | 15.59% (±5.93) | 4.54% (±0.75) | 11.69% (±2.6) | 19.23% (±6.23) | 7.91% (±1.45) | 11.55% (±3.01) | 4.82% (±0.2) | Volume maximum total displacement |
| **TotVonMises** | 15.51% (±2.91) | 7.85% (±0.31) | 9.42% (±0.46) | 12.17% (±1.72) | 19.49% (±6.71) | 16.73% (±5.38) | 8.70% (±0.56) | Volume maximum total Von Mises stress |
| **CurlDisp_Y*** | 7.32% (±0.61) | 5.44% (±1.88) | 9.13% (±2.92) | 7.93% (±1.74) | 9.35% (±2.58) | 7.30% (±1.30) | 7.39% (±4.72) | Volume maximum curl displacement in Y-direction |
| **PrincStrain_X*** | 6.41% (±0.59) | 5.07% (±0.12) | 3.54% (±0.94) | 3.94% (±0.97) | 5.97% (±0.18) | 3.34% (±0.32) | 5.70% (±0.15) | Volume maximum principal strain in X-direction |
| $f_1$ | 2.41% (±0.45) | 1.63% (±0.04) | 2.04% (±0.1) | 2.07% (±0.06) | 2.23% (±0.15) | 2.15% (±0.26) | 2.22% (±0.08) | First eigenfrequency |
| $f_2$ | 2.61% (±0.62) | 1.49% (±0.08) | 2.12% (±0.07) | 2.04% (±0.12) | 1.97% (±0.09) | 2.03% (±0.24) | 1.92% (±0.05) | Second eigenfrequency |
| $f_3$ | 2.00% (±0.13) | 1.43% (±0.01) | 1.93% (±0.04) | 3.39% (±2.11) | 2.17% (±0.42) | 1.84% (±0.07) | 1.93% (±0.06) | Third eigenfrequency |
| $[f_1, f_2, f_3]$ | 6.56% (±0.33) | 4.93% (±0.07) | 6.48% (±0.26) | 6.90% (±0.94) | 6.58% (±0.20) | 6.33% (±0.37) | 6.26% (±0.14) | First three eigenfrequencies |
| **npf1_RMS** | 0.44% (±0.14) | 0.17% (±0.07) | 1.12% (±0.91) | 0.71% (±0.38) | 0.60% (±0.04) | 0.44% (±0.25) | 0.32% (±0.08) | Root Mean Squared normalized participation factor of the first eigenfrequency |
| **npf2_RMS** | 0.67% (±0.32) | 0.22% (±0.03) | 0.36% (±0.05) | 1.37% (±1.02) | 0.36% (±0.05) | 1.06% (±0.49) | 0.21% (±0.07) | Root Mean Squared normalized participation factor of the second eigenfrequency |
| **npf3_RMS** | 9.95% (±2.37) | 6.31% (±1.60) | 0.66% (±0.25) | 2.95% (±3.43) | 7.90% (±1.69) | 0.32% (±0.08) | 8.17% (±0.66) | Root Mean Squared normalized participation factor of the third eigenfrequency |
| **npf123_RMS** | 15.26% (±2.65) | 6.21% (±0.79) | 1.66% (±0.64) | 5.04% (±3.60) | 17.37% (±5.69) | 4.28% (±4.79) | 8.8% (±0.19) | Root Mean Squared normalized participation factor of the first three eigenfrequencies |

* label values were scaled before training

## APPENDIX F
### TRAINING GRAPHS

When training models, a dynamic stopping condition was adopted whereby the training was aborted if the validation loss had not improved after 50% of the maximum number of epochs that the algorithm should be run. The model that performed best on the validation data was saved at each epoch. If the model failed to satisfy the stopping condition after the maximum number of epochs had been reached, the number of training epochs was doubled. This approach ensured that stalled training sessions would be aborted whereas those that yielded improvements would continue. We also generated training logs showing the model being trained past the point of divergence between the training loss and the validation loss. Since the best model was saved at each epoch, the training graph shows the training past the point where the best model was saved.

TABLE IV
PERFORMANCE OF MODEL USED FOR GEOMETRIC OPTIMIZATION OF BEAM CROSS-SECTIONS

| dataset | $f_1$ MAE | $f_1$ MAPE | $f_2$ MAE | $f_2$ MAPE | $f_3$ MAE | $f_3$ MAPE | Avg MAE | Avg MAPE |
|---|---|---|---|---|---|---|---|---|
| test | 2.8674 | 2.0664 | 3.5548 | 2.2883 | 14.0474 | 1.7375 | 6.8232 | 2.0307 |
| train | 2.1925 | 1.5608 | 2.7394 | 1.7755 | 10.6596 | 1.3132 | 5.1972 | 1.5498 |
| validation | 2.8848 | 2.0644 | 3.5594 | 2.3033 | 14.0507 | 1.7294 | 6.8316 | 2.0323 |

*A. Training Graph of Model Used For Geometric Optimization of Beam Cross-Sections*

Please see table IV for performance figures, and refer to the training graph in figure 2.

## APPENDIX G
### DATASETS

Our datasets are hosted on Mendeley Data. Access the following DOI to download the datasets: 10.17632/y3m8xm6kfk.

## APPENDIX H
### DATA GENERATION

For each dataset, we set the beam extrusion length, spikiness, irregularity, and the maximum volume from pixel errors ($VPE$).

3

Fig. 2. Training graph of the model used for geometric optimization of beam cross-sections. We selected the model with the lowest validation error during training.

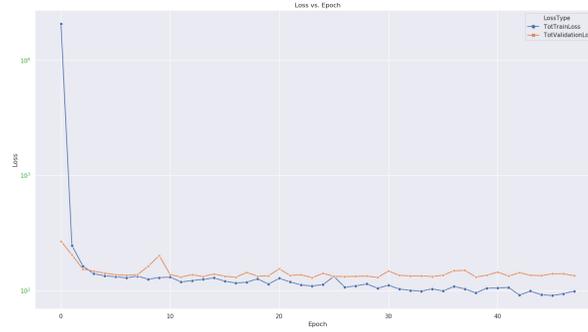

$$VPE = \frac{abs(Volume - NRPX * L)}{Volume} \qquad (1)$$

The $VPE$ is a measure of how well the cross-section image captures the actual beam shape. The $VPE$ is the absolute percentage error between the volume determined from pixels and the actual beam volume computed by FreeCAD (see eqn. 1). Volume is computed from pixels by multiplying the number of pixels that make up the beam cross-section in the image ($NRPX$) by the beam length ($L$). All pixels in a 128×128 pixel image are counted, whereby each pixel is scaled to the 1 $mm^2$ size. By eliminating shapes that are outside of the $VPE$ threshold, we ensured that the cross-section images adequately represent the actual beam shape and thereby partially controlled the noise level in our dataset.

Please visit our cantilever beam dataset generator GitHub repository (https://github.com/ResearchMedia/CantileverBeamDatasetGenerator) to access the code and instructions to reproduce our datasets or generate new ones.

## APPENDIX I
## COMSOL MULTIPHYSICS: DATA ANALYSIS SCRIPTS

Please visit our cantilever beam dataset generator GitHub repository (https://github.com/ResearchMedia/CantileverBeamDatasetGenerator) to download the COMSOL Multiphysics applications used for the FEA static and frequency analyses in this work.

## REFERENCES

[1] ARORA, R., BASU, A., MIANJY, P., AND MUKHERJEE, A. Understanding deep neural networks with rectified linear units. In *6th International Conference on Learning Representations, ICLR 2018 - Conference Track Proceedings* (2018).
4

17



\end{document}